# On the Nature of Measurement in Quantum Mechanics

DOUGLAS M. SNYDER
LOS ANGELES, CALIFORNIA

## Abstract

A number of issues related to measurement show that self-consistency is lacking in quantum mechanics as this theory has been generally understood. Each issue is presented as a point in this paper. Each point can be resolved by incorporating a cognitive component in quantum mechanics. Measurement in quantum mechanics involves the *meaning* of the physical circumstances of the experiment. This meaning is in part independent of what traditionally are considered purely physical considerations.

## Text

The following five points concern issues in quantum mechanics that indicate that experimenter variables play a role in this theory. Each point is followed by a brief discussion.

### POINT ONE

1.  An experimenter oftentimes may consider a physical interaction *either* as a measurement involving a *macroscopic* measuring instrument interacting with an observable *or* in terms of a set of *microscopic* systems, including the observable, functioning lawfully in accordance with the Schrödinger equation.

    The question arises: On the microscopic level, does the abrupt change in the wave function associated with the observable that often occurs in a measurement occur when the observable and the macroscopic apparatus are considered as a set of microscopic physical systems interacting with each other? According to Bohr, the answer apparently is, "No." Bohr emphasized that it was the macroscopic nature of the measuring apparatus that was responsible for a measurement. According to Bohr (1935), there was some loss of control due to the macroscopic character of the physical apparatus used to make a measurement. This lack of control resulted in the limitation on knowledge described by the uncertainty principle. Thus if the macroscopic system is considered in terms of its microscopic components, a measurement should not

occur. There should not be an abrupt change in the wave function associated with the observable.

Here is how Bohr considered measurement in quantum mechanics. He wrote:

> It is imperative to realize that in every account of physical experience one must describe both experimental conditions and observations by the same means of communication as one used in classical physics. In the analysis of single atomic particles [in quantum mechanics], this is made possible by *irreversible amplification effects* [emphasis added]–such as a spot on a photographic plate left by the impact of an electron, or an electric discharge created in a counter device–and the observations concern only where and when the particle is registered on the plate or its energy on arrival at the counter. (Bohr, 1955/1958b, p. 88)

Bohr (1955/1958a) essentially defined the *phenomenon* in quantum mechanics as the entire experimental context, specifically the experiment that is conducted on an observable. Bohr maintained that the experiment conducted was described in terms of the physical apparatus used and the observable measured and that the physical world in quantum mechanics could only be understood in terms of the phenomenon.

> On the lines of objective description, it is indeed more appropriate to use the word phenomenon to refer only to observations obtained under circumstances whose description includes an account of the whole experimental arrangement. In such terminology, the observational problem in quantum physics is deprived of any special intricacy and we are, moreover, directly reminded that every atomic phenomenon is closed in the sense that its observation is based on registrations obtained by means of suitable amplification devices with irreversible functioning such as, for example, permanent marks on a photographic plate, caused by the penetration of electrons into the emulsion. (Bohr, 1955/1958a, p. 73)[1]

---

[1] Citing Bohr, Wheeler (1981/1983) wrote: "A phenomenon is not yet a phenomenon until it has been brought to a close by an irreversible act of amplification such as the blackening of a grain of silver bromide emulsion or the triggering of a photodetector....We are dealing with an

# On the Nature

An experiment discussed by Feynman, Leighton, and Sands (1965) that will be discussed shortly indicates that the destruction of interference often associated with a measurement in quantum mechanics does not depend on a macroscopic measuring instrument.

POINTS TWO AND THREE

2. A measurement can occur on a *microscopic* level without depending on a macroscopic measuring instrument.

3. The presence or absence of interference may depend on a *comparison* of different observables in an experiment, even if the observables cannot actually be observed by the experimenter.

*The Importance of Comparison*

In demonstrating the point that measurement does not rely on a macroscopic physical apparatus acting as a measuring instrument interacting with the observable measured, Feynman *et al*. considered the destruction of interference on the atomic level in the case of neutron scattering by a crystal. This destruction of interference depends on a change in the nucleus in the crystal that interacts with the impacting neutron allowing the experimenter to "in principle, find out which nucleus had done the scattering, since it would be the only one with spin turned over" (Feynman *et al*., 1965, chap. 3, p. 8). Initially, all the nuclei in the crystal have their spin components along a specific spatial axis set in one direction. It is a change in the spin component along a spatial axis for a nucleus in the crystal impacted by a neutron that distinguishes this nucleus from other nuclei for which the spin component along this same axis has not changed and which thus all remain in the original direction.

Note that the destruction in interference depends on the nuclei that are not impacted by the neutron all having spin components that are *opposite* in orientation to the nucleus that is impacted by the incoming neutron. It is not the physical interaction between a nucleus in the crystal and the neutron that alone is responsible for the destruction of interference. If, in the interaction, the spin component for the nucleus does not flip and the spin component for the neutron therefore does not flip, interference is not destroyed. In the experiment, destruction of the interference depends on a *microscopic* event, a spin component flip on the part of the neutron that *distinguishes* the path of the

---

event that makes itself known by an irreversible act of amplification, by an indelible record, an act of registration." (pp. 184-185, 196)

impacting neutron from alternative paths. The different spin components cannot, of course, be observed by the experimenter.

It is the information conveyed to the observer that determines whether or not interference is destroyed. The physical interaction alone does not account for the different physical results. Feynman *et al*. (1965) had a different view:

> Well, if we can tell which atom [nucleus] did the scattering, what have the other atoms [nuclei] to do with it? Nothing of course. The scattering is exactly the same as that from a single atom. (chap. 3, p. 8)

When Feynman *et al*. wrote: "Nothing of course," their own description of the experiment indicates their statement is incorrect. Perhaps, they could argue that on a physical level alone the other nuclei are not involved in the scattering, or that if considered in isolation the impacted nucleus would not show interference. Their own description of the destruction of interference, though, in the crystal depends on the *difference* between the spin component orientations of the impacted nucleus and the other nuclei. In other words, the experimental circumstance involving numerous nuclei is not the same as one involving only one nucleus.

### POINT FOUR

4. A measurement needs to make which-way information only *in principle* available to destroy interference.

In different works on quantum theory, Feynman distinguished those circumstances in which wave amplitudes associated with an observable are first added and the absolute value of the sum is squared to derive the probability of an event, as opposed to taking the absolute square of each of the wave amplitudes and taking their sum to determine the probability of the same event. Feynman *et al*. (1965) wrote concerning quantum mechanics:

> [1] The probability of an event in an ideal experiment is given by the square of the absolute value of a complex number $\phi$ which is called the probability amplitude...

> [2] When an event can occur in several alternative ways, the probability amplitude for the event is the sum of the probability amplitudes for each way considered separately...

> [3] If an experiment is performed which is capable of determining whether one or another alternative is actually taken, the probability of the event is the sum of the probabilities for each alternative. (chap. 3, p. 10)

In their experiment discussed above, Feynman *et al*. (1965) wrote that the distinction between the different paths need not actually be known to an observer.[2] All that is necessary is that their distinction is in principle possible in the experimental circumstances.

> If you could, *in principle*, distinguish the alternative *final* states [of the observables] (even though you do not bother to do so), the total, final probability is obtained by calculating the *probability* for each state (not the amplitude) and then adding them together. If you *cannot* distinguish the final states *even in principle*, then the probability amplitudes must be summed before taking the absolute square to find the actual probability. (Feynman *et al.*, 1965, chap. 3, p. 9)

Point four might appear to indicate that the observer is not central to measurement in quantum mechanics. It is important to note that distinguishing final paths constitutes a measurement. Remember, oftentimes, the physical interaction occasioning a measurement can be a macroscopic apparatus interacting with a microscopic observable *or* it can be an interaction among microscopic observables alone (Point one). Because of these different *views* one can take of an interaction, the *in principle* possibility of distinguishing paths supports a cognitive component to measurement in quantum mechanics.

In this context, it is important to note negative observation. In a negative observation, a measurement does not involve a physical interaction at all (Bergquist, Hulet, Itano, & Wineland, 1986; Cook, 1990; Epstein, 1945; Nagourney, Sandberg, & Dehmelt, 1986; Renninger, 1960; Sauter, Neuhauser, Blatt, & Toschek, 1986; Snyder, 1996, 1997). A negative observation occurs where an observation is made by deducing that a particular physical event must have occurred because another physical event did not occur with subsequent consequences for the functioning of the physical world stemming from the change in knowledge.

---

[2] In *QED: The Strange Theory of Light and Matter*, Feynman (1985) maintained the same principles hold in quantum electrodynamics. In sum, where one can distinguish between the possible routes, or developmental sequence, for an observable, interference between the component wave amplitudes characterizing the existent is destroyed.

POINT FIVE

5.  Macroscopic devices may *either* : 1) interact with an observable acting as a measuring instrument where coherence is destroyed and there is an abrupt change in the wave function associated with the observable, *or* 2) act as a device in the service of separating a wave function into coherent components or recombining coherent component wave functions so that interference is expressed. There is nothing in the nature of a macroscopic device itself that necessarily signals a measurement.

An example is a double-hole diaphragm that is rigidly attached to a support through which electrons pass on their way to a screen that they impact and that records their position (Bohr, 1949/1969). Coherent component wave functions are developed as a result of the electron passing through the diaphragm, resulting in interference which is detectable when the electrons impact the screen. A macroscopic apparatus in this case does not result in a measurement when it interacts with a microscopic observable. Something else is needed to distinguish whether a measurement is made.

If, instead of being fixed, the diaphragm was movable due to momentum transfer between the electron and the diaphragm, a position measurement would occur. Which hole an electron went through on its way to the screen could be determined by how the diaphragm moved as the electron passed through (Bohr, 1949/1969). What is it about allowing the diaphragm to move that changes its function so dramatically? The answer lies in the destruction of interference, and interference in quantum mechanics depends on coherent waves that are mathematically complex and do not have a physical existence.

One can in fact piece together coherent component wave functions associated with an observable using a macroscopic physical apparatus. For example, one may use a one-half silvered mirror to recombine coherent component wave functions associated with a photon that had originally been separated earlier by another one-half silvered mirror. The macroscopic apparatus used to combine the component wave functions does not destroy interference. It actually implements interference using the coherent wave functions.

CONCLUSION

Following is a summary of the points made in this paper.

1. In many experimental circumstances, a macroscopic apparatus appears to be responsible for the occurrence of measurements in quantum mechanics.

2. Feynman *et al.*'s experiment works as they discussed, and their position that a macroscopic physical apparatus is not always necessary for a measurement is sound.

3. The occurrence of a measurement may depend on a *comparison* of observables, even if these observables cannot be observed but can only be thought about.

4. A physical interaction that is considered a measurement, accompanied by an abrupt change of the wave function associated with the observable measured, does not necessarily have to be considered as such. The collapse of the wave function need not occur in another *view* of the same interaction that focuses on the interacting systems microscopically.

5. A measurement can be said to occur when *in principle* it is possible to distinguish possible measurement outcomes, even if these outcomes are not actually measured.

6. Negative observations are possible in quantum mechanics.

7. A macroscopic physical apparatus is not necessary for a measurement, and indeed it can be used to develop interference.

Notice the use of the terms *view*, *comparison*, and *in principle*. The inconsistencies noted in this paper indicate that quantum mechanics is not concerned exclusively with the physical. A measurement affecting an observable can be made without a physical interaction. Quantum mechanics, though, always does depend on the mathematically complex waves associated with observables, and interference itself combines these complex waves. Anything that can be known about an observable is derived from these waves.

Ascribing a cognitive component to quantum mechanics accounts for the terms indicating cognition that are noted above as well as the result stemming from Bohr's concept of measurement that depending on one's view there may or not be an abrupt change of the wave function associated with an observable when a physical interaction occurs. One can summarize the points in this paper by characterizing measurement in quantum mechanics as involving the *meaning* of the physical circumstances of the experiment. This meaning is in part independent of what traditionally are considered purely physical considerations.